\documentclass[twocolumn,floatfix]{revtex4}
\usepackage{graphicx}
\usepackage{color}
\usepackage{array}
\usepackage{longtable}

\begin{document}
\title{Do scientists trace hot topics?}
\author{Tian Wei$^{1}$}
\author{Menghui Li$^{2,3}$}
\author{Chensheng Wu$^{1,4}$}
\author{XiaoYong Yan$^{1,5}$}
\author{Ying Fan$^{1}$}
\author{Zengru Di$^{1}$}
\author{Jinshan Wu$^{1,\ast}$}
\affiliation{
1. Department of Systems Science, School of Management, Beijing Normal University, Beijing, 100875, P.R. China \\
2. Temasek Laboratories, National University of Singapore, 117508, Singapore \\
3. Beijing-Hong Kong-Singapore Joint Centre for Nonlinear $\&$ Complex Systems (Singapore),
National University of Singapore - Kent Ridge, 119260, Singapore \\
4. Beijing Institute of Science and Technology Intelligence, Beijing, 100044, P.R China \\
5. Centre for Complex Systems Research, Shijiazhuang Tiedao University, Shijiazhuang, 050043, P.R. China
}

\begin{abstract}
Do scientists follow hot topics in their scientific investigations?
In this paper, by performing analysis to papers published in the
American Physical Society (APS) Physical Review journals, it is
found that papers are more likely to be attracted by hot fields,
where the hotness of a field is measured by the number of papers
belonging to the field. This indicates that scientists
generally do follow hot topics. However, there are qualitative
differences among scientists from various countries, among research
works regarding different number of authors, different number of
affiliations and different number of references. These observations
could be valuable for policy makers when deciding research funding
and also for individual researchers when searching for scientific
projects.

\noindent \\
\noindent{\bf SUBJECT AREAS:} Statistical Physics, Scientometric, Complex Networks
\end{abstract}

\maketitle

The phenomenon of ``the rich get richer'', which is also called
preferential attachment in the field of complex networks
\cite{BAmodel}, is quite common in many fields \cite{Simon, Price}
(for example, see references cited in Table \ref{tablealpha});
however, the scientific field is composed of scientists, a special
group of people who focus on proposing, investigating and
implementing original and creative ideas. Therefore, it is plausible
that the ``the rich get richer'' phenomenon is less pronounced in
the fields investigated by scientists than in other areas. Ideally
scientists choose their fields of investigation according to their
scientific interest and the scientific value of the investigated
question but not due to the hotness of the investigated fields.  In
this work, using published papers from the American Physical Society
(APS) Physical Review journals beginning in $1976$ and ending in
$2009$, we test whether the subject of a new paper is more likely to
be in a hot field than in a relatively unknown field when the paper
is published. We also compare scientists from different countries.
Such comparisons could provide insightful and interesting
information. In China, modern scientific development is still very
young. It is believed among many scientists that there are many more
Chinese scientists that are followers than original thinkers
compared with many other countries. In this work, we offer direct
empirical support for this hypothesis. Finally, we also determine if
the degree of tracing hot fields differs for papers with different
number of authors or affiliations and different number of
references. Interestingly, it is found that scientists who
collaborate with more authors or more affiliations tend to follow
hotter fields than those who works with a few collaborators or
affiliations. Moreover, papers with a small number of references, on
average, are more interest-driven or value-driven, whereas papers
with a large number of references are more hotness-driven. These
empirical discoveries, particularly if it is also performed in other
fields and for a larger periods of time, could provide valuable
information for policy makers.

\noindent\\
\textbf{Results}
\noindent\\
Empirically, the phenomenon of preferential attachment
has been found in many systems. We compile a list of typical
systems, where their exponent $\alpha$ values are shown in Table
\ref{tablealpha}. Later, we will compare our results on hotness
tracing of newly published papers against other phenomena listed in
this table.

\begin{figure*}
\includegraphics[width=0.8\linewidth]{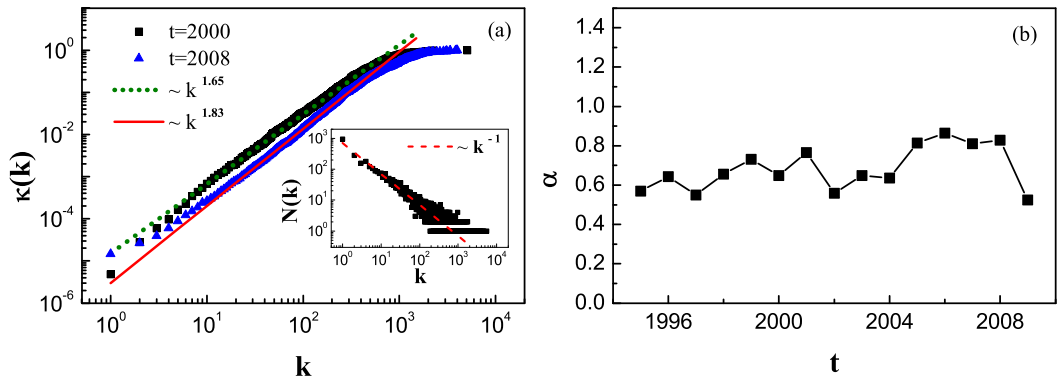}
\caption{\label{overall} Empirical preferential attachment to hot
fields of new papers in the PR-PACS data set. (a) The cumulative
probability functions $\kappa (k)$ in the years 2000 and 2008.
Inset: the distribution function of the sizes of fields. (b) The
exponents $\alpha$ for different years.  We start our
measurements from 1995, not 1976, the beginning year of the data, since the method requires a relatively large
initial system. Notice $\alpha=0.52-0.87$ is near the lower end of
all the exponents collected in Table \ref{tablealpha}. The 2009 exponent is relatively smaller compared to its previous several years for
reason that we do not yet know.  If not mentioned, $\Delta t$ is one
year in this paper. The straight lines are guide to the eye through
this paper.}
\end{figure*}

Firstly, we examine the phenomenon of preferential attachment of
papers in the PR-PACS data set (see \emph{Methods}
for details). In a log-log plot, Fig. \ref{overall} (a) displays the
accumulated distribution function $\kappa\left(k\right)$ with
respect to the size $k$ of the field that a newly published paper
belongs to. The positive exponents $\alpha$ indicate that new papers
are more likely to focus on hot fields (larger sizes). Or to say,
generally, scientists do publish more new papers in current hot
fields. We obtain the exponents $\alpha$ by least-square fitting
from $k=1$ to $k=300$ as the curves deviate from the straight line
for large $k$ due to low statistics. For different years $t$, all of
$\kappa\left(k\right)$ follow power law, namely
$\kappa\left(k\right) \sim k^{\alpha +1}$, but slightly different
parameters $\alpha$ (as shown in Fig. \ref{overall} (b)). We also
plot the distribution of fields' size $N(k)$ as an inset in Fig.
\ref{overall} (a), which follows a highly skewed distribution.

Compared with the preferential attachment phenomena in other fields
listed in Table \ref{tablealpha}, the exponents $\alpha=0.52 - 0.87$
from our PR-PACS data are near the lower end of all the exponents.
Therefore, although as we pointed out already overall scientists do
publish more on hot topics, scientific works do appear to be more
independent and more interest-driven or value-driven than other
fields. Out of all the other $12$ entries in Table \ref{tablealpha},
only sexual contact in sexual networks ($\alpha=0.32-0.80$)
\cite{sexual} , users attached to membership of groups of Digg
($\alpha=0.69$) \cite{group} and  friendship relations in Linkedin
($\alpha=0.6$) \cite{Answers} are approximately at the same level of
hotness tracing. It might be easy to ``follow'' a star member in a
social networking website, such as Flickr ($\alpha=1.0$)
\cite{Answers}; however, it might not be so easy to shift a research
field, join a user group of a different product or change sexual
partners towards hotter choices.

\begin{figure*}
\includegraphics[width=0.8\linewidth]{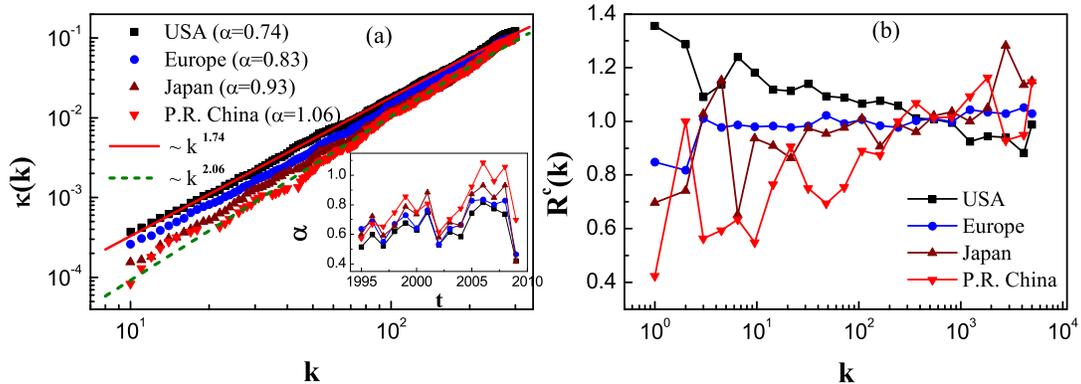}
\caption{\label{countries_overall} Results for authors from the
most-contributed countries in the PR-PACS data set. (a) The
cumulative probability functions $\kappa (k)$ in the year 2008 for
USA, Europe, Japan and P. R. China. The exponent of Chinese authors
($\alpha=1.06$) is much larger than that of other countries. Inset:
The exponents $\alpha$ for different years, where the exponents
$\alpha$ of China always are larger than that of USA. (b) The
relative ratio $R^{c}(k)$ for USA, Europe, Japan and P. R. China.}
\end{figure*}

Secondly, we test whether the intensity of tracking hot topics of
scientific research differs in different countries or regions.
Therefore, we classify the papers according to countries (region) of
the first author's affiliation, and calculate the absolute
contribution ratios $r^{c}$ of several major countries
(region) within PR-PACS data set. They are USA ($33.07\%$), Europe
($39.0\%$), Japan ($6.94\%$) and China  ($3.73\%$). As seen from
Fig. \ref{countries_overall} (a), in the year 2008 the
exponent ($\alpha = 1.06$) of P. R. China is larger than that of
other countries (region), e.g., USA $0.74$, Europe $0.83$ and Japan
$0.93$. Moreover, as shown in the inset of Fig.
\ref{countries_overall} (a), the exponent $\alpha$ for China is
generally larger than that of other countries (region) for different
years. These results indicate that the phenomenon of tracing hotness
is more severe among Chinese scholars.

To provide a comparative study, we also calculate the relative
contribution ratio $R^{c}(k)$ of papers from country $c$ and
belonging to the fields with size $k$ (see \emph{Methods} for the
details). As seen from Fig. \ref{countries_overall} (b), the
relative contribution ratio $R^{c}(k)$ of Chinese scholars is
smaller than $1$  in cold fields (small $k$) but larger
than $1$ in hot fields (lager $k$), indicating that Chinese scholars
make less contributions to cold fields than their average
contribution but more to hot fields than their average. Meanwhile, the situation of USA
is opposite to that of China. This difference also indicates that
Chinese scientists are more keen to follow hot topics than United
States scientists from another aspect. This agrees with our previous
observations.

Considering the fact that scientific studies in China are still
young, it is understandable that a large percentage of them are on
hotness-driven fields rather than value-driven fields. The
discovered order -- the USA, Europe, Japan and China -- of degree of
hotness tracing makes sense intuitively. These results are more or
less consistent with our intuitions. How different positions are
related to scientific policies of that country, or even the culture
and values of that country, although is definitely worth a further
investigation, is outside the scope of the current study. We simply
want to demonstrate the capability of the methods that are discussed
above in analyzing publication records, and to present some basic
discoveries using the methods in this work.

\begin{figure}
\includegraphics[width=0.8\linewidth]{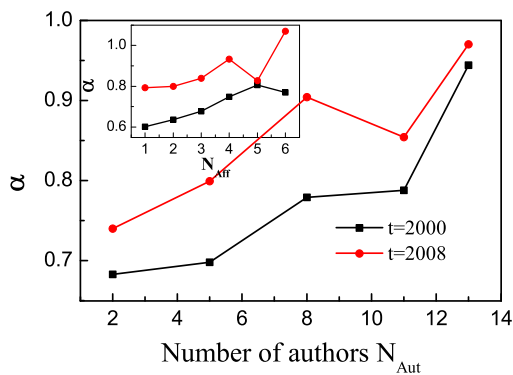}
\caption{\label{authoraffiliation} The preferential attachment
exponents $\alpha$ with respect to number of authors $N_{Aut}$
(number of affiliations $N_{Aff}$ in the inset) in the year 2000 and
2008. Note that every three numbers for authors are
grouped together and labelled as the intermediate number, e.g., 1,2
and 3 are grouped together and labelled as 2. Overall, the exponent
$\alpha$ increases with number of authors and affiliations.}
\end{figure}

Next, we measure the influence of different number of authors and
affiliations on the degree of tracing hot topics. Therefore, we
classify the papers according to their number of authors and number
of affiliations. It is argued in Ref. \cite{Nature} that research
works with many authors or many affiliations typically focus more on
hot topics because it might exactly be the hotness of the paper
subject that made collaboration attractive among the scientists and
that a joint task team is generally more likely to focus on
short-term projects rather than long-term projects. Here we make
such an examination based on the PR-PACS data. We can see from Fig.
\ref{authoraffiliation} that overall, $\alpha$, the degree of
severity of tracing hot topics, increases with the number of authors
and affiliations. These results provide empirical supports for
the arguments in Ref. \cite{Nature}. In a sense, global
collaboration is not necessarily a good strategy for high-quality,
value-driven research topics as suggested in Ref. \cite{Nature}.

\begin{figure}
\includegraphics[width=0.8\linewidth]{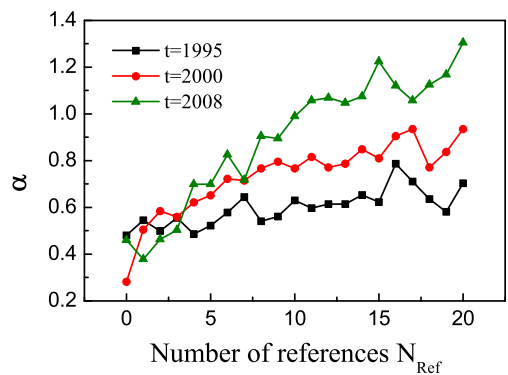}
\caption{\label{references} Preferential attachment exponent
$\alpha$ with respect to the number of references $N_{Ref}$.
Clearly, the exponent $\alpha$ increases with number of references.
The absolute value of $\alpha$ is a bit larger in recent years than
in early years for larger number of references. }
\end{figure}

Finally, we investigate the effect of the number of references on
tracking hot topics. We classify the papers according to their
number of references. It is obvious that the average number of
references in papers today is much larger than that of early times. For
earlier times, one can intuitively hypothesize that a pioneer paper
or a paper of good quality typically cited less references. However,
today, the number of references may or may not relate to how
innovative the paper is. Here, we examine this hypothesis. As seen
from Fig. \ref{references}, the exponents $\alpha$, the degree of
severity of tracing hot topics, increase with the
number of references, which could indicate that papers with a
larger number of references are more likely to be on hot topics.
Notice that the absolute values of $\alpha$ for larger number of
references in recent years are larger than that in earlier years,
suggesting that scientists trace hot fields a bit more severely in
recent years than in earlier times. Therefore, the hypothesis is
reasonable overall.

\noindent \\
\textbf{Discussion}
\noindent \\
We have found that in the tested fields of science for papers
published by the APS physical review journals, hot fields attract
more newly published papers; however, scientific works are not as
hotness-driven compared with other fields. Among the major
countries, this phenomenon of tracing hotness occurs more in China
compared with other countries, which could be related to the fact
that China is still a developing country in terms of scientific
research.
We also found that papers with more authors,
more affiliations and more references were on subjects that were
more hotness-driven. This observation could potentially be valuable
to policy makers that fund scientific projects.

Here only data of physics publications were analyzed. A cross-field
comparison would be interesting, particularly if one can do similar
studies on math and social science, such as economics. Similar
studies can be applied on producing, selling or buying products.
That is to investigate when a product is manufactured,
sold or bought, how often is it related to the number of the product
that has been manufactured, sold or bought already. Such studies
could be valuable for marketing research.

It might also be interesting to determine how often
outstanding/important works when they are first published are in hot
fields. Fields gradually or suddenly become hot after major
breakthroughs are presented in a few pioneer papers, which later on,
might be awarded or honoured. Using records of awards such as the
Nobel prize, the Fields Medal, the Turing Award {etc.}, it would be
possible to perform a statistical analysis of papers with major
breakthroughs. All the award-winning papers could be collected,
where the same analysis to determine $\alpha$ can be performed,
which could then be compared with the overall $\alpha$.

The above investigation can be performed, not only at the macro
level of counties of authorship but also at the mesoscopic level of
affiliations of authorship. In this way, one might be able to
compare tradition, culture and research quality at various levels of
academic units. In principle, one could also collect all published
papers of one author and treat all those papers as a subset and
study the preferential attachment phenomenon of those papers if the
size of such collection is large enough. It would then be possible
to use it to measure the creativity and originality of a researcher.

\noindent \\
\textbf{Methods}
\noindent \\
\textbf{Data description and notations.} Our data set is a
collection of all papers published by the American Physical Society
(APS) Physical Review journals from $1976$ to $2009$. Each paper is
recorded as a data entry, which includes its title, date of
publication, classification number according to the AIP-Physics and
Astronomy Classification Scheme (PACS), author(s), affiliation(s)
and reference(s) to other papers within the data set. The entire
data set contains more than $320,000$ entries, including a variety
of article types, such as article, brief report, rapid
communication, comment, reply, erratum, essay, announcement,
editorial, announcement and so on. Here, we will only consider those
research papers, e.g., article, brief report and rapid
communication, with PACS numbers and refer to it as the PR-PACS data
set. At last, we have $M=315,082$ entries, which each entry,
{\it{\i.e.}} each paper, is denoted as $i$, and $N=5,472$ PACS,
which each entry, {\it{\i.e.}} each PACS number, is denoted as
$\lambda$ .

We use PACS, the established hierarchical classification systems of
physics, to identify the fields within the physics field.
Mathematically, we can use one matrix to characterize the relation
between paper and PACS. $A$ is an $M\times N$ adjacency matrix, with
element $a_{i\lambda}= 1$ if paper $i$ belongs to PACS $\lambda$,
otherwise 0. We define the size of a field, $k_{\lambda}=\sum_{i}
a_{i\lambda}$, as the number of papers that belong to it. Then, we
can calculate the number $N(k)$ of fields with size $k$.

\textbf{Measuring preferential attachment.} Here, we explain our
method for this statistical analysis for testing preferential
attachment on temporal data set. We calculate the empirical value of
the relative probability $T(k)$ that a new paper published within a
short period $\Delta t$ connects to a field which has a size of $k$
before the time $t$ \cite{Medline} as follows. Since the
corresponding time-dependent absolute probability $P_{k}(t)$ that a
new paper published in a field with size $k$ is proportional to
$T(k)n_{k}(t)/N(t)$, where $n_{k}\left(t\right)$ is the number of
fields with size $k$ and $N(t)$ is the number of fields immediately
before time $t$, then $T(k)$ can be estimated by making a histogram
of the sizes $k$ of the fields to which each paper is added within
the time period $\Delta t$ in which each sample is weighted by a
factor of $\frac{N(t)}{n_{k}(t)}$,
\begin{equation}
T(k) = \sum_{i,\lambda}^{k_{\lambda}(t)=k}{\frac{a_{i\lambda}N(t)}{n_{k}(t)}}
\end{equation}
where $k_{\lambda}(t)=k$ means that the field, to which
the papers published within the period $\Delta t$ belong, has
size $k$ at time $t$. We now have the empirical curve $T(k)$
from the above statistical analysis. In order to conveniently
compare $T(k)$ with different time $t$, $T(k)$ is normalized as
$T'(k) = \frac{T(k)}{\sum_{k'} T(k')}$ \cite{Citation, measurePA}.

The preferential attachment hypothesis states that the rate $T'(k)$
with which a node with $k$ links acquires new links is a
monotonically increasing function of $k$ \cite{BAmodel}, namely
\begin{equation}
T'(k) = \frac{k_{i}^{\alpha}}{\sum_{j}{k_{j}^{\alpha}}} =
C(t)k_{i}^{\alpha}
\end{equation}
For BA model $\alpha = 1$ \cite{BAmodel}. To obtain a smooth curve
from noisy data, we take the cumulative function form instead of
$T'(k)$:
\begin{equation}
\kappa\left(k\right) = \int_0^k T'(k)\,dk^{'}
\end{equation}
Thus, $\kappa\left(k\right)$ should be proportional to $k^{\alpha +
1}$. We can now fit the empirical curve from the previous
statistical analysis and then compare it against this hypothesized
curve of preferential attachment. This is the general procedure of
all the analysis presented in this work.

To test the preferential attachment of scientific research differs
in different countries, we separated the entire data set according
to countries of the first author's affiliation and then perform a
comparison among the most contributed countries or regions (USA,
Europe, Japan, China). With this separated data set, we perform the
examination of preferential attachment only counting the papers from
authors in country $c$ as
\begin{equation}
T^{c}(k) = \sum_{i,\lambda}^{k_{\lambda}(t)=k, aff_{i}=c}{\frac{a_{i\lambda}N(t)}{n_{k}(t)}}.
\end{equation}
Here $aff_{i}=c$ means the principle affiliation of this paper $i$ is in country $c$.
In counting $k$ and $n_{k}(t)$, we included papers from all
countries, meaning that scientists from all countries face the
temptation of tracing the same overall hotness in the entire PR-PACS
data set. Similarly, besides countries, the above calculation can be applied to any
features of papers, such as different number of authors,
affiliations and references.

\textbf{Measuring relative contribution ratio
$R^{c}\left(k\right)$.} Absolute contribution from a country is
measured simply by a percentage of published papers from that
country out of the total number of published papers,
\begin{equation}
r^{c} = \frac{\sum_{k} m^c_k}{\sum_{k} m_k},
\end{equation}
where $m_k$ ($m^c_k$ ) is the number of papers (from
country $c$) belonging to fields with size $k$. Here we present a
more detailed breakdown of this absolute contribution by looking at
each individual field what is the percentage of papers from that
country out of all papers in that field, and then normalized by the
absolute contribution of that country,
\begin{equation}
R^{c}(k)= \frac{m^c_k}{m_k}\frac{1}{r^{c}}.
\end{equation}
This is a static measure, so it is easy to perform. In a sense it
also describes how often scientists in that country are pursuing hot
fields.

\noindent\\
{\bf Acknowledgements} This work was supported by NSFC Grant
$11205014$ and $60974084$. The authors thank the APS Physical Review
for sharing the data. There has been many very fruitful discussions
between one of us (J. Wu) and Prof. Bertrand Roehner.

\noindent\\
{\bf Author Contributions Statement} J. W., Z. D. and Y.F. designed
the research. T. W., M. L., X. Y. and C. W. analyzed the data. T.W.
and J. Wu participated in the writing of the manuscript.

\noindent\\
{\bf Additional Information}\\
 Competing interests statement: The
authors declare that they have no competing financial interests.

\noindent\\
Correspondence should be addressed to J. Wu (jinshanw@bnu.edu.cn).

\begin{table}
\caption{A list of some values of $\alpha$, the degree of
preferential attachment collected from literature, showing also the
number of nodes $N$, the number of links $E$. The first column is
the name of the investigated database, and  the discussed
relationships are discussed within the brackets.}
\begin{tabular}{lrccc}\hline
 Network& $N$ & $E$ &$\alpha$ & Ref. \\
\hline
 \color{blue}{APS-PACS} \color{blue}{(belonging)}&  \color{blue}{5,472}& \color{blue}{900,832} & \color{blue}0.52 - 0.87 & \color{blue}{$-$}\\
Sexual networks\\
(sexual contact) & 260 - 1,220 &$-$& 0.32 - 0.80 &\cite{sexual}\\
Digg\\
(Membership of group) &212,635&1,185,167 &0.69&\cite{group}\\
Linkedin (friendship) &7,550,955 &30,682,028 & $0.6$ &\cite{Answers}\\
\hline
Medline (coauthorship)& 1,648,660&$-$&$1.04\pm0.04$ &\cite{Medline}\\
NYGI(coproduction) & 10,000&700,000& $1.20\pm0.06$ &\cite{NYGI}\\
Bar (communication) & 3,988&$-$&$1.25\pm 0.13$ &\cite{Loco}\\
Google(communication) & 39,918&$-$&$1.36\pm 0.14$ &\cite{Loco}\\
Flickr (following)  &584,207 &3,554,130& $1.0$ &\cite{Answers}\\
aNobii\\
(friendship, following)&86,800  &697,910& $1.0$ &\cite{aNobii}\\
Douban (following)&1,614,288  &14,573,170& $0.95$ &\cite{Douban}\\
Wealink (friendship) & 223,482&273,209& $1.0$ &\cite{Wealink}\\
Citation (Citation) & 1,736&83,252& $0.95\pm0.1$ &\cite{Citation}\\
\hline
\end{tabular}\label{tablealpha}
\end{table}

\end{document}